\documentclass[11pt]{article}
\usepackage{a4}
\usepackage{times}
\usepackage{graphicx}
\usepackage{color}

\begin{document}
\title{Tuning Tempered Transitions}
\author{Gundula Behrens\thanks{Department of Epidemiology and Preventive Medicine, University Hospital Regensburg
93042 Regensburg, Germany; Gundula.Behrens@klinik.uni-regensburg.de},
Nial Friel\thanks{School of Mathematical Sciences,
University College Dublin, Belfield, Dublin 4, Republic of Ireland; 
Email: nial.friel@ucd.ie}
and Merrilee Hurn\thanks{Department of Mathematical Sciences, 
University of Bath, Bath, BA2 7AY, UK; 
Email: M.A.Hurn@bath.ac.uk}
}

\maketitle
\begin{abstract}
\noindent
The method of tempered transitions was proposed by Neal (1996) for
tackling the difficulties arising when using Markov chain Monte Carlo to sample
from multimodal distributions.
In common with methods such as simulated tempering and Metropolis-coupled
MCMC, the key idea is to utilise a series of successively easier to sample
distributions to improve movement around the state space.
Tempered transitions does this by incorporating moves through
these less modal distributions into the MCMC proposals.
Unfortunately the improved movement between modes comes at a high 
computational cost with a low acceptance rate of expensive proposals.
We consider how the algorithm may be tuned to
increase the acceptance rates for a given number of temperatures.
We find that the commonly assumed geometric spacing of temperatures is
reasonable in many but not all applications.

\end{abstract}

\section{Introduction to tempering ideas in MCMC}
It is well known that standard Markov chain Monte Carlo (MCMC) methods,
such as the Metropolis-Hastings algorithm or the Gibbs sampler, often
have difficulties in moving around their target distribution.
When a chain mixes poorly in this way, there is a danger that modes have
been missed or that modes are not represented in their right proportions,
both of which may lead to bias in the statistical inference. 
To overcome such mixing problems, various more sophisticated MCMC 
methods have been devised based on a few key ideas.
This paper concentrates on one of these key ideas, namely tempering.

One way to motivate tempering is to think of using importance
sampling to estimate some expectation
$\mathsf{E}_{p_0}[h(X)]$ with respect to the 
target distribution $p_0$ by sampling from some less modal distribution 
$p_1$.
One possibility for generating a less modal distribution than $p_0$ on the same
support is to ``flatten'' it by taking
$p_1 (x) \propto p_0(x) ^\beta, \ \forall x$,
with $\beta < 1$.
As $\beta \rightarrow 0$, $p_1 (x)$ becomes closer to a uniform distribution
and consequently becomes more amenable to sampling.
For $\beta$ close to 1, there is far less benefit as $p_1$ may not be that much
less modal than $p_0$.
Unfortunately as the two distributions become far enough apart that the
difficulty with modality is overcome, they may also become far enough apart so that
many of the importance weights will be very close to zero resulting in
unstable estimates of $\mathsf{E}_{p_0}[h(X)]$.
The basis of all the tempering methods is the introduction of
a series of distributions 
bridging the gap between $p_0$ and $p_1$.
The differences between the various approaches is in how these
bridging distributions are included.
We will describe various approaches to incorporating bridging
distributions using the common form of tempering which involves powering
up all or part of the unnormalised target distribution.
The inclusion of other types of bridging distribution would also be possible,
but the literature has generally restricted itself to this form.

We assume that the target distribution can be written as 
\begin{equation} 
\label{eq:target}
p(x) \propto \pi(x)\, \exp( -\beta_0 \, h(x) ),
\end{equation}
where $h(x)$ may be known as the ``energy'' function and the parameter
$\beta_0$ as the target ``inverse temperature''. 
Since we can write any positive function $f(x)$ in exponential form
$f(x)= \exp (- \beta_0 \, h(x) )$ by setting
$h(x)=- \frac{1}{\beta_0} \log (f(x))$ this 
class covers a wide range of applications. 
The tempered distributions are then defined by
\begin{equation} 
\label{eq:tempered}
p_{i}(x) \propto \pi(x)\, \exp( -\beta_i \, h(x) ),
\qquad i=0,1,\ldots,n,
\end{equation}
where $0 \le \beta_n \le \ldots \le \beta_1 \le \beta_0$, are the inverse
temperatures characterising each distribution.
The flexibility of potentially only tempering part of the target distribution
is quite useful.
In Bayesian problems it may be that one or other of the
prior and likelihood contribute to the mixing problems.

One of the earlier suggestions for incorporating tempering
into MCMC is to run $n+1$ Markov chains in parallel, each sampling from one
of the $n+1$ tempered distributions.
At each iteration, proposals are made to update each chain separately and
additionally there is a proposal to swap the $x$ values between chains thereby
coupling them and giving rise to the name Metropolis-coupled
MCMC (Geyer 1991\nocite{GeyerML}).
The state space is the enlarged set of $(n+1)$ values for $x$ and the
target distribution is $p_0 \otimes p_1 \otimes \ldots \otimes p_n$.
The idea is that large moves made under $p_n$ will filter back down
to the lowest level $p_0$.
The normalising constants for the tempered distributions are not needed
in this method as they appear only in the acceptance probabilities for the
coupling move where they cancel out.
However the tempered distributions do need
to be close in order that the swaps between them are not too infrequent.
This may mean that $n$ will have to be large and there are then
obvious consequences for storage and computational effort.

A single chain alternative to Metropolis coupling is simulated tempering 
(Marinari and Parisi 1992\nocite{MarinariSimulated}, 
Geyer and Thompson 1995\nocite{GeyerAnnealing}) 
which runs a chain on the state space of $x$ augmented by a
variable $i$ which takes the values $i=0,1,\ldots,n$ with probabilities
determined by a ``pseudo-prior''. 
The stationary distribution of $x|i$ is $p_{i}(x)$, and updates 
are either of $x|i$ or $i|x$ with the latter effectively moving up or down
the tempering sequence.
Although again the normalising constants of the tempering distributions 
are not needed explicitly, in practice to get reasonable acceptance rates for
the moves between temperatures, the pseudo-prior needs to be
roughly proportional to these unknown normalising constants. 

Tempered transitions (Neal 1996\nocite{NealTempered}) is another single chain
method but without the need to guesstimate the
relative normalising constants of the tempered distributions.
It uses a deterministically ordered sweep up and down the tempering
distributions as a way of generating proposals for
the main chain in a way which will be described in more detail in the next
section.
The overwhelming cost of the algorithm is in the construction of the 
proposals and therefore it is imperative that these are tuned carefully to
maximise acceptance rates.
Neal (1996\nocite{NealTempered}) finds tempered transitions and simulated
tempering to be of roughly equal computational cost.
He also compares tempered transitions, simulated tempering and
Metropolis-coupled MCMC on other criteria such as storage requirements
and the number of tempering levels required concluding that there is
no overall winner and that the choice of method may be problem and goal 
specific.

Closely related methods which aim to make fuller use of the sampling at
all temperature levels via importance sampling can be found in
Neal (2001)\nocite{NealAnnealedImportance} and more recently in Gramacy,
Samworth and King (2010\nocite{GramacyImportance}).
The former has links to tempered transitions, effectively using just the
second half of the complex proposal mechanism.
The latter has stronger connections with simulated tempering and Metropolis
coupled MCMC where samples from the different temperatures are stored.
Other instances of ideas involving populations of samples can be found in 
both the 
population-based MCMC and the Sequential Monte Carlo literature (see
Jasra, Stephens and Holmes (2007)\nocite{JasraPopulationSimulation} for an
overview).

A common question arising across the algorithms involving tempering is
the choice of the bridging distributions given by Equation~(\ref{eq:tempered}).
The general recommendation is to space the $\beta$s geometrically,
that is so that $\beta_i/\beta_{i+1}$ is a constant for all levels $i$
(Neal 1996\nocite{NealTempered}).
Neal formulates this rule by considering sampling from a multivariate Gaussian
using simulated tempering; the geometric spacing attempts to maximise 
the acceptance rates of swaps between neighbouring chains at all levels.
Gelman and Meng (1998)\nocite{Gel98} also consider choices of bridging 
distribution, although in the context of the closely related question
of estimating normalising constants where they are trying to minimise
the Monte Carlo error of path sampling estimates.
Other work on rationales for choosing the $\beta$s can be found in
Iba (2001)\nocite{Iba01} and Lefebvre, Steele and Vandal (2010)\nocite{Lef08}.
The former reviews the (largely physics) literature, comparing simulated
tempering with exchange and ensemble Monte Carlo methods and aims to
maximise the swapping rates between the bridging distributions 
using preliminary runs (to satisfy a theoretically derived optimality
criterion). 
The latter is interested in path sampling for estimating normalising constants
and the related tuning of the bridging distributions; they derive an
expression for the symmetrised Kullback-Leibler divergence between pairs of
distributions and use the minimisation of this as their criterion.

In this paper we consider tempered transitions with bridging distributions
of the form given by Equation~(\ref{eq:tempered}).
Of the various tempering schemes, the choice of the $\{\beta_i\}$ seems to have
been least well addressed for tempered transitions.
Our approach is largely computational and tries to answer the question 
``For a given number $n$ of tempering distributions, how best should they be spaced?''. 
We note that the question of how we should choose $n$, for fixed 
computational time, is beyond the scope of this article.
The approach we take is to use a small number of preliminary short runs to assess 
whether geometric spacing is likely to be adequate and, if not, we propose an optimal way 
of spacing them. In Section \ref{sec:efficiency} we describe tempered transitions in
detail, building on many of the insights offered in Neal's paper. We provide
a theoretical analysis which outlines when geometric temperature spacing is optimal
and give a motivating example where geometric spacing is sub-optimal. We also 
draw some parallels with some of the other theoretical approaches outlined above.
In Section \ref{sec:applications} we discuss the implementation details
of applying our proposed approach to a slowly mixing MCMC application.

\section{How to tune tempered transitions?}
\label{sec:efficiency}

\subsection{The tempered transitions algorithm}
We begin by describing the algorithmic details of the tempered transitions
algorithm and setting up the reasoning behind our tuning approach.
Suppose the chain is currently in state $x$, then
the algorithm generates a proposal $x^\prime$ for the
next state using a secondary chain which passes through all
the auxiliary distributions $\left\{p_{i} \right\}$ first in
ascending order of the $\beta$s (``heating-up'') and then in descending order
(``cooling-down'') back to the target distribution
$p_{0} $.
To do this, it uses $n$ pairs of MCMC transition kernels with the $i^{th}$
pair, $T_i$ and $T^\prime_i$ satisfying detailed balance with respect to
$p_{i}$
\begin{displaymath}
p_{i}(x)\, T_{i}(x,x^{\prime})=
p_{i}(x^{\prime})\,T^\prime_{i}(x^{\prime},x)\ \ \ 
\forall x, x^{\prime}, i=1,\ldots,n.
\end{displaymath}

\begin{description}

\item[Step 1] Set $x_0=x$.

\item[Step 2] Move up and down the tempered distributions using MCMC transitions

Generate $x_1$ from $x_0$ using transition kernel $T_{1}$.\\ 
Generate $x_2$ from $x_1$ using transition kernel $T_{2}$.\\
$\vdots$\\
Generate $x_n$ from $x_{n-1}$ using transition kernel $T_{n}$.\\
Generate $x_{n-1}^{\prime}$ from $x_n$ using transition kernel $T^\prime_{n}$.\\ 
$\vdots$\\
Generate $x_1^{\prime}$ from $x_2^{\prime}$ using transition
kernel $T^\prime_{2}$.\\ 
Generate $x_0^{\prime}$ from $x_1^{\prime}$ using transition
kernel $T^\prime_{1}$. 
 
\item[Step 3] Accept $x^\prime = x_0^{\prime}$ as the next state with probability
\begin{equation} 
\alpha( x , x^\prime | {x}_0 , {x}_1 , \ldots , {x}_n , x_{n-1}^{\prime}, \ldots, x_1^{\prime } , x_0^{\prime} )
= \min \left\{ 1, 
\left[ \prod_{i=0}^{n-1}
\frac{p_{{i+1}}(x_i)}{p_{i}(x_i)}
\right]
\, \left[ \prod_{i=0}^{n-1}
\frac{p_{{i}}(x_i^{\prime})}{p_{{i+1}}(x_i^{\prime})}
\right]
\right\},
\label{eqn:accept}
\end{equation}
otherwise, remain at state $x$.
\end{description}
There is no need for the normalising constants 
of the tempering distributions as they cancel in the acceptance probability.
Neal (1996\nocite{NealTempered}) demonstrates that the algorithm satisfies
detailed balance with respect to the target $p_{0}$.
However it is perhaps clear that the proposal is potentially 
computationally costly and tricky to tune.

There are two dependent aspects to the tuning.
The first is that the $\{ T_i , T_i^\prime \}$ should have reasonable
acceptance rates and, at the later stages of the tempering, be
able to make large moves in the state space
(otherwise this expensive proposal makes little change).
This is not quite the usual tuning problem for MCMC in that 
each successive level has a different target distribution.
During the first half, these distributions are becoming 
progressively less modal, while in the second half the reverse is true.
Obviously the closer the consecutive distributions, ie the closer the
consecutive $\beta$s, the less of an effect this will be.

\begin{figure}[t] 
\begin{center}
\includegraphics[height=12.5cm, angle=270]{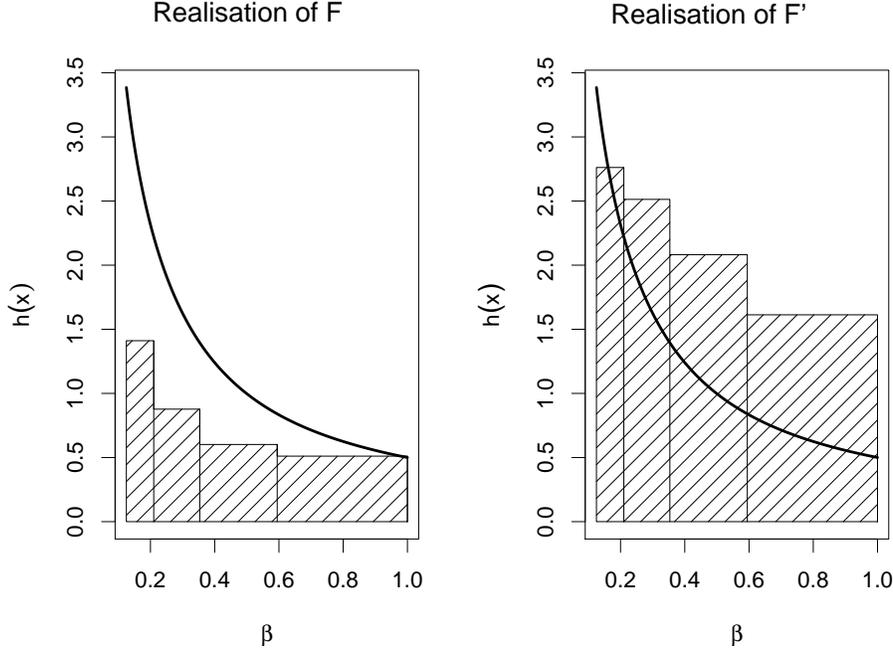}
\caption{Two approximations to the integral of $g(\beta)$.
The breakpoints of the rectangles are given by 
$\beta_n \le \ldots \le \beta_{1} \le \beta_0$.
The shaded area on the left is
$F= \sum_{i=0}^{n-1} (\beta_{i}-\beta_{i+1})h(x_{i})$, while that on
the right is 
$F^\prime=\sum_{i=0}^{n-1} (\beta_{i}-\beta_{i+1})h(x_i^\prime)$.
The overlaid curve is $g(\beta)=\mathsf{E}_\beta [h(X)]$.}
\label{fig:fig2}
\end{center}
\end{figure}

Subject to the individual $\{ T_i , T_i^\prime \}$ working well, 
the second tuning aspect is that the overall acceptance rate of the
entire tempered transition proposal should be as high as possible (although
notice that if all the proposed changes in the tempering are rejected, the
overall proposal will be accepted since $x_0 = x_1 = \ldots = x_1^\prime =
x_0^\prime$ and so a high acceptance rate here can be slightly
misleading if viewed in isolation). 
To gain more insight into tuning the tempered transition acceptance rate,
we follow Neal's
lead and rewrite the acceptance probability, Equation (\ref{eqn:accept}),
using the form of tempering defined by Equation (\ref{eq:tempered}).
\begin{eqnarray} 
\alpha( x , x^\prime | {x}_0 , {x}_1 , \ldots , x_1^{\prime } , x_0^{\prime} )
& = & \min \left\{ 1, 
\left[ \prod_{i=0}^{n-1}
\frac{\exp(-\beta_{i+1}h(x_i))}{\exp(-\beta_{i}h(x_i))}
\right]
\, \left[ \prod_{i=0}^{n-1}
\frac{\exp(-\beta_{i}h(x_i^{\prime}))}{\exp(-\beta_{i+1}h(x_i^{\prime}))}
\right]
\right\} \nonumber \\
\label{eqn:accept2}
& = & \min \left\{ 1,
\exp(-(F^\prime-F))
\right\} \\
\mbox{where } F & = & \sum_{i=0} ^ {n-1} (\beta_i-\beta_{i+1})h(x_i) \nonumber \\
\mbox{and } F^\prime & = & \sum_{i=0} ^ {n-1} (\beta_{i}-\beta_{i+1})h(x_i^\prime). \nonumber
\end{eqnarray}
This expression has an interpretation related to estimating the ratio of
normalising constants by thermodynamic integration. 
Let $Z(\beta)$ denote the normalising constant of the distribution defined by
Equation (\ref{eq:tempered}), where for the moment we assume that $\beta$ takes
continuous values in the interval $[\beta_0,\beta_n]$.
\begin{eqnarray}
Z( \beta ) & = & \int_x \pi(x) \exp( -\beta h(x) )\ dx \nonumber \\
\mbox{then }\ \ \ \frac{d Z( \beta )}{d\beta} & = & \int_x \pi(x) 
\frac{d\exp( -\beta h(x) )}{d\beta}\ dx \nonumber \\
& = & \int_x (-h(x))\frac{\pi(x) 
\exp( -\beta h(x) ) }{Z( \beta )}Z( \beta )\ dx \nonumber \\
& = & -Z( \beta ) \mathsf{E}_\beta [h(X)]. 
\label{eqn:thermo2}
\end{eqnarray}
Solving this differential equation for $Z(\beta)$ gives the relation
\begin{equation}
\log( Z(\beta_n)) - \log( Z(\beta_0)) = \int_{\beta_n}^{\beta_0}
\mathsf{E}_\beta [h(X)] \ d\beta.
\label{eqn:thermo}
\end{equation}
That is, the log of the ratio of the normalising constants at two values
of $\beta$ can be expressed as the area under the curve
$g(\beta)=\mathsf{E}_\beta [h(X)]$ between them.
Recall that the sequences $\{x_0,\ldots x_{n-1}\}$ and 
$\{x_{n-1}^\prime,\ldots x_{0}^\prime\}$ are drawn such that the $x_i$
have target distribution $p_i$, while the $x_i^\prime$ have target distribution
$p_{i+1}$.
Figure \ref{fig:fig2} illustrates a slightly idealised
realisation of $F$ (left) and
$F^\prime$ (right) as the shaded areas constructed as the sum of
rectangles with widths $(\beta_{i}-\beta_{i+1})$ and heights 
$h(x_i)$ (left) or $h(x_i ^\prime)$ (right).
Both areas are approximations of the integral of $g(\beta)$ between
$\beta_n$ and $\beta_0$.
Different realisations of $\{x_0,x_1,\ldots,x_0^\prime\}$ will obviously
give quite different and usually rather messier pictures, 
with correspondingly quite varied values of $F^\prime - F$.
(In fact, Figure \ref{fig:fig2} was constructed using the average of
several realisations to reduce this variability for presentation
purposes.)
Those realisations of $x_0,x_1,\ldots,x_1^\prime,x_0^\prime$ 
where the shaded area on the right ($F^\prime$) is smaller than that on the
left ($F$) will be accepted since in that case $\exp(-( F^\prime - F))> 1$ 
in Equation (\ref{eqn:accept2}).
Those for which $F^\prime$ is slightly larger than $F$ may be accepted,
but we will almost certainly reject those for which $( F^\prime - F)$ is 
large.
We take this as a motivation for selecting the $\{\beta_i\}$ for fixed $n$.

\subsection{The proposed rationale for choosing $\{\beta_i\}$}
Given the cost of each tempered transition proposal, our motivation is to
increase the number of proposals accepted. 
The value of $\beta_0$ is fixed by Equation (\ref{eq:target}), and we 
assume that the other extreme of the $\beta$s is also determined, possibly
by the fact that it defines a distribution for which direct sampling is
possible, certainly by the need to move around the state space freely under
$p_n$.
What remains undetermined are $n$ and the set
$\{\beta_1, \ldots, \beta_{n-1}\}$.

\begin{figure}[t] 
\begin{center}
\includegraphics[height=12.5cm, angle=270]{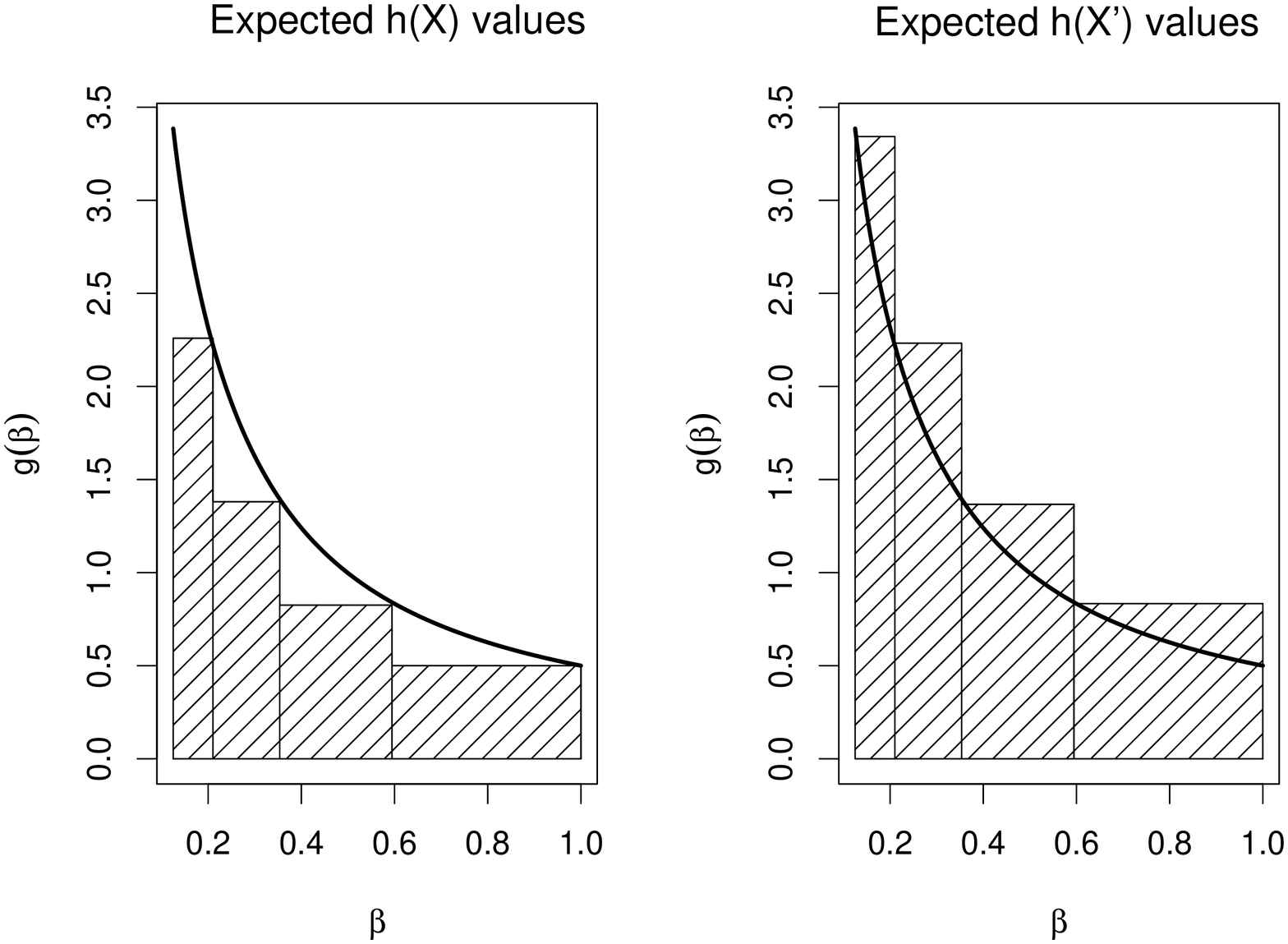}
\caption{
The shaded area on the left is
$\sum_{i=0}^{n-1} (\beta_{i}-\beta_{i+1})g(\beta_i)$, while that on
the right is 
$\sum_{i=0}^{n-1} (\beta_{i}-\beta_{i+1})g(\beta_{i+1})$.
The difference between the two areas is $S_n(\beta_0,\ldots,\beta_n)$.}
\label{fig:fig1}
\end{center}
\end{figure}

Figure \ref{fig:fig2} showed the $F$ and $F^\prime$ associated with a 
realisation $\{x_0,x_1,\ldots,x_0^\prime\}$.
If at each stage the transitions were able to reach their equilibrium
distributions in the one step available, 
then $\mathsf{E}[h(x_i)]=g(\beta_i)$ and
$\mathsf{E}[h(x_i^\prime)]=g(\beta_{i+1})$;
Figure \ref{fig:fig1} shows the corresponding
approximations to the integral of $g(\beta)$ (this is equivalent to
Neal's Figure 1(a)).
Denote this difference between the areas of these two step functions
as a function of the $\beta$ values
\begin{eqnarray}
S_n(\beta_0,\ldots,\beta_n) & = & 
\sum_{i=0} ^ {n-1} (\beta_{i}-\beta_{i+1})\mathsf{E}_{i+1}[h(X)] -
 \sum_{i=0} ^ {n-1} (\beta_{i}-\beta_{i+1})\mathsf{E}_{i}[h(X)] \nonumber \\
& = & \sum_{i=0} ^ {n-1} (\beta_{i}-\beta_{i+1})g(\beta_{i+1}) -
 \sum_{i=0} ^ {n-1} (\beta_{i}-\beta_{i+1})g(\beta_i). 
\label{eqn:sums}
\end{eqnarray}
Some results are well known for $g(\beta) = \mathsf{E}_\beta [h(X)]$.
Rewriting Equation (\ref{eq:tempered}) as
\begin{equation}
p_\beta (x) = \pi(x)\, \exp( -\beta \, h(x) - K( \beta) ) 
\end{equation}
so that $K(\beta)$ is the log of the normalising constant $Z(\beta)$
and rearranging Equation (\ref{eqn:thermo2}),
\begin{eqnarray}
g(\beta )
& = & - \frac{ 1}{ Z( \beta )} \frac{d Z( \beta )}{d\beta} \nonumber \\
& = & - \frac{d \log( Z( \beta )) }{d\beta} \nonumber \\
& = & - K^\prime ( \beta )
\end{eqnarray}
\begin{eqnarray}
\mbox{ and } g ^\prime ( \beta ) & = & 
\int h(x) \frac{d}{d \beta} p_{\beta}(x) dx \nonumber\\
& = & \int h(x) ( -h(x) - K^\prime(\beta) ) \pi(x)
\exp( -\beta \, h(x) - K( \beta) ) dx \nonumber\\
& = & \int (-h(x)^2 +h(x)g(\beta) ) p_{\beta}(x) dx \nonumber\\
&= & - \mbox{Var}_{\beta}\left[ h(X) \right].
\label{eqn:gprime}
\end{eqnarray}
Therefore $g ^\prime ( \beta ) <0$, for all $\beta$, showing that $g(\beta)$ is 
a decreasing function of $\beta$. 
It is possible to examine $g ^{\prime\prime} (\beta)$ similarly, 
showing that the curve may be convex, concave, or a mixture of the two.
The main point here is that because $g(\beta)$ is decreasing, we know that
$S_n(\beta_0,\ldots,\beta_n) \ge 0$.

We propose the minimisation of $S_n(\beta_0,\ldots,\beta_n)$ over $\{\beta_i\}$ as 
our rule for choosing the tempered transition parameters.
Obviously increasing $n$ immediately reduces $S_n$. 
However our primary
motivation here is the most effective choice of the particular
$\{\beta_1,\ldots,\beta_{n-1}$ for a fixed number of levels
$n$ and fixed values of $\beta_0$ and $\beta_n$.

Note that minimising $S_n = \mathsf{E}[ F^\prime-F]$ is not 
directly equivalent to 
maximising the expected value of the acceptance probability, 
$\alpha = \min\{ 1 , \exp ( - (F^\prime-F))\}$, however 
\begin{equation}
\mathsf{E} [ \exp ( - (F^\prime-F)) ] = 1 - S_n + 
\frac{\mathsf{E} [ (F^\prime-F)^2]}{2!} - \ldots 
\end{equation}
and so intuitively minimising $S_n$ seems a reasonable start.
Other possible criteria include, for example, maximising 
$\mathsf{P} ( F^\prime < F)$ over the $\{\beta_i\}$ or,
as suggested by one of the referees, examining the variance as
well as the expectation of 
$F^\prime-F$ since a high variance could perhaps improve mixing by
generating big moves more often than a low variance might.
We have so far only considered looking at $S_n$.

\subsection{A motivating example}
\label{sec:witch}
To motivate the tuning procedure, we study the one-dimensional
two-parameter simplified Witch's Hat distribution used by Geyer and Thompson
(1995\nocite{GeyerAnnealing}).
Although this is quite a straightforward example, we shall
see that it is one for which geometric spacing of the temperatures is not
optimal.
Geyer and Thompson attribute this distribution to Matthews (1993)
\nocite{Mat93} who introduced it as a problem case for the Gibbs sampler:
\begin{displaymath}
\label{eqn:witches}
p( x ) \propto 1 + b \mathsf{I}_{[x \le a]},\ \ \ 0 \le x \le 1
\end{displaymath}
where the parameters satisfy $0 < a < 1$ and $b \ge 0$.
This apparently innocuous L-shaped distribution causes problems for standard 
Metropolis-Hastings moves if $a$ is small but $b$ is large as it
can be difficult to move between the intervals $[0,a]$ and $(a,1]$.
The distribution can be expressed in a form suitable for tempering 
\begin{displaymath}
p_i( x ) \propto \exp( \beta_i 
\log(1 + b \mathsf{I}_{[x \le a]}) ) ,\ \ \ i = 0, \ldots, n
\end{displaymath}
where $\beta_0 = 1$.
In this case, $g(\beta) = \mathsf{E}_\beta 
[- \log(1 + b \mathsf{I}_{[x \le a]}) ]$ is available analytically
as are its derivatives:
\begin{eqnarray}
g( \beta ) & = &
\frac{-a(1+b)^{\beta}\log(1+b)}
{a(1+b)^{\beta}+(1-a)} \nonumber \\
g^\prime( \beta ) & = & 
\frac{a(a-1)(1+b)^{\beta} (\log(1+b)) ^2}
{(a(1+b)^{\beta}+1-a)^2} \nonumber \\
g^{\prime\prime}( \beta ) & = & 
\frac{-a(a-1)(1+b)^{\beta} (\log(1+b))^3}
{\left( a(1+b)^{\beta}+(1-a)\right) ^3}
\left( a(1+b)^{\beta}-(1-a) \right)
\end{eqnarray}
The second derivative shows that for $\beta \in [0,1]$, the
curve $g(\beta)$ may be convex
(if $a \in [0.5,1]$ and $b \ge 0$), concave (if $a \in (0,0.5)$ and
$b \in (0,1/a-2]$), or a 
mix of the two (otherwise).
We propose to study one distribution for which $g(\beta)$ is convex (taking
$a=0.5$ and $b=7.5 \times 10^8$) and one for which it is concave (taking
$a=10^{-4}$ and $b=9.5 \times 10^3$).
The latter distribution is hard to sample with roughly half the mass
concentrated in the narrow $[0,9.5 \times 10^3]$ peak.
The former distribution does not present a sampling problem,
but we are still interested in the effect of the shape of $g(\beta)$ on 
tempering performance.

\begin{figure}[t] 
\begin{center}
\includegraphics[height=13.25cm,width=16cm]{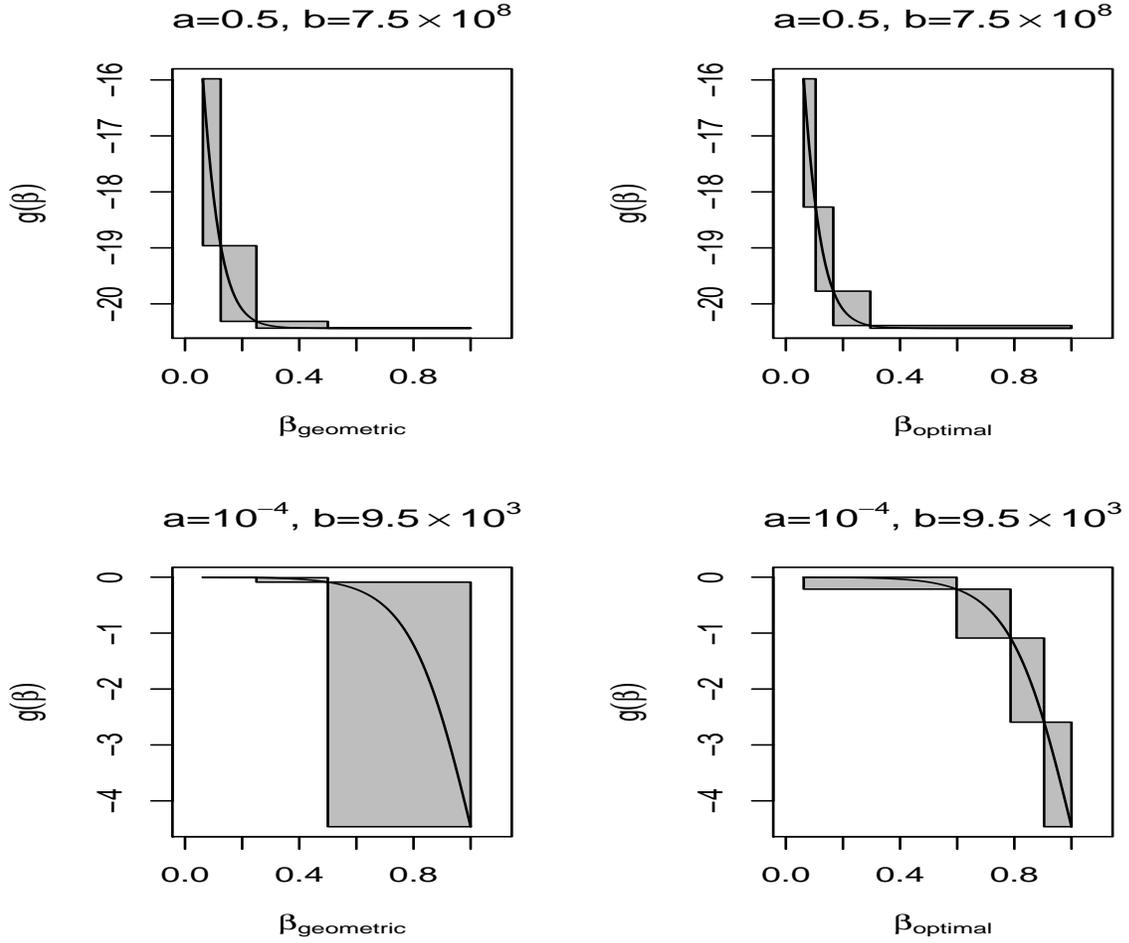}
\caption{
$S_n$ for the Witch's hat distribution when $n=4$ using geometric $\{\beta_i\}$
spacing (left) or the optimal $\{\beta_i\}$ (right)
overlaid by $g(\beta)$ in black;
on the top row, $a=0.5$ and $b=7.5 \times 10^8$, on the bottom row
$a=10^{-4}$ and $b=9.5 \times 10^3$.}
\label{fig:minsn}
\end{center}
\end{figure}

Given $g(\beta)$, $\beta_0$, $\beta_n$ and $n$, how do we minimise $S_n$
over $\{\beta_1,\ldots,\beta_{n-1}\}$, 
Equation (\ref{eqn:sums})?
The $(n-1)$ partial derivatives are available,
\begin{equation}
\frac{ \partial S_n }
{ \partial \beta_i} = (g(\beta_{i-1})-2g(\beta_i)+g(\beta_{i+1}))
+ (\beta_{i-1}-2\beta_i+\beta_{i+1})g^\prime (\beta_i) ,\ \ i=1,\ldots,n-1,
\label{eqn:derivs2}
\end{equation}
however, despite their relatively simple form, no analytic solution 
is readily available for the minimisation problem.
As a result, we perform the minimisation numerically using the built-in
quasi-Newton \textsf{optim} with option \textsf{L-BFGS-B} in \textsf{R} 
incorporating derivative information and the constraints
that $\beta_n < \beta_i < \beta_0$ for $i=1,\ldots,n-1$ and fixed $\beta_0$
and $\beta_n$.
This function is not guaranteed to converge to a global maximum;
in our experiments it was insensitive to starting 
points (we used geometric or uniform spacings) with the exception of 
occasional catastrophic convergence for large $n$ to a non-ordered 
set of $\beta$s.
In all cases we encountered, changing from geometric to uniform
initial spacings, or vice versa, resolved this problem.
Figure \ref{fig:minsn} illustrates the minimal $S_n$ and the corresponding
$\{ \beta_i \}$ for the two examples of the Witch's hat distribution when
$n=4$ and $\beta_n=1/16$, as
well as the equivalent $S_n$ for a geometric spacing of the $\{ \beta_i \}$.
Unsurprisingly given the different shapes of the two curves,
the change in 
the size of $S_n$ achieved by the optimal scheme over the geometric one is
more significant for the concave $g(\beta)$ curve.
However, even in the convex example it is clear that the values of the $\beta$s
themselves, particularly $\beta_1$, are quite different under geometric spacing and the minimal $S_n$ scheme.
Table \ref{tab:witch}
shows the minimum values of $S_n$ and the geometric values of $S_n$
for $n=2$, 4, 8, 16, 32 and $64$ and for both pairs of parameters $a$ and $b$.
For each $n$ we performed 500000 iterations of tempered transitions where
at each level $i = 1, \ldots, n$ we use direct sampling (by inversion) to
draw from the tempered distribution $p_i$.
This direct sampling is only realistically possible, at least for values of $\beta$ close to $\beta_0$, in a test example such as this, but it does allow us 
to separate the effects of the different $\beta$ choices
from the effects of slow mixing of the transitions at levels 1 to $n$.
Table \ref{tab:witch} 
gives observed average acceptance rates together with the estimated integrated 
autocorrelation time of the tempering calculated with respect to the known 
theoretical mean.
The Witch's Hat example is unusual in that only
moves between the regions $0 \le x \le a$ and $a < x \le 1$ are 
problematic while all within-region moves are always accepted (giving rise
to unusually high acceptance rates for a tempering problem).
In addition, a high acceptance rate in tempered transitions can actually mask
a lack of mixing and so integrated autocorrelation times are a useful
diagnostic.

\begin{table}[t]
\begin{center}
\begin{tabular}{|l|c|c|c|c|c|c|c|}
\hline
& & $n=2$ & $n=4$ & $n=8$ & $n=16$ & $n=32$ & $n=64$ \\
\hline \hline
& Optimal & & & & & & \\
$a=0.5$ & $S_{n}$ & 0.83386 & 0.30241 & 0.13214 & 0.06218 & 0.03023 & 0.01492 \\
$b=7.5 \times 10^8$ & $\alpha$ & 0.78 & 0.80 & 0.84 & 0.87 & 0.91 & 0.93 \\
(convex)& $\tau$ & 1.55 &1.48 & 1.38 & 1.28 & 1.20 & 1.14 \\
& Geometric & & & & & & \\
& $S_{n}$ & 0.90444 & 0.38612 & 0.18454 & 0.09122 & 0.04548 & 0.02272 \\
& $\alpha$ & 0.78 & 0.79 & 0.82 & 0.85 & 0.89 & 0.89 \\
& $\tau$ & 1.58 & 1.51 & 1.46 & 1.36 & 1.26 & 1.26 \\
\hline \hline
& Optimal & & & & & & \\
$a=10^{-4}$ & $S_{n}$ & 1.46627 & 0.63456 & 0.29879 & 0.14591 & 0.07234 & 0.03607 \\
$b=9.5 \times 10^3$ & $\alpha$ & 0.55 & 0.63 & 0.72 & 0.80 & 0.85 & 0.90 \\
(concave)& $\tau$ & 7.05 & 2.36 & 1.75 & 1.47 & 1.33 & 1.22 \\
& Geometric & & & & & & \\
& $S_{n}$ & 3.34158 & 2.20779 & 1.25229 & 0.64996 & 0.32786 & 0.16428 \\
& $\alpha$ & 0.51 & 0.51 & 0.55 & 0.61 & 0.69 & 0.78 \\
& $\tau$ & 591.36 & 55.56 & 9.13 & 3.11 & 1.91 & 1.54 \\
\hline
\end{tabular}
\caption{Results for the Witch's Hat problem under two 
settings of the parameters $a$ and $b$ and multiple choices of $n$, the
number of tempering levels.
The minimal sum of squares, observed acceptance rates and estimated integrated
autocorrelation times are shown for the geometric scheme and for the optimal scheme.}
\label{tab:witch}
\end{center}
\end{table}

For both distributions, decreasing $S_n$ by
optimising the $\{\beta_i\}$, increases the observed acceptance rate and
decreases the integrated autocorrelation time.
These improvements are most noticeable when
comparing the geometric and optimal schemes for the hard sampling
problem, $a=10^{-4}$ and $b=9.5 \times 10^3$, where the changes in $S_n$
are most dramatic.
Concentrating on $n=4$ to tie in with Figure \ref{fig:minsn}, for the
easier convex $g(\beta)$, optimising the
$\{\beta_i\}$ made a small difference to the overall $S_n$ albeit with
noticeable changes to the $\{\beta_i\}$ themselves.
Here the tuning has made only marginal improvements in acceptance rates
and integrated autocorrelation times (although as noted earlier, this 
distribution is not hard to sample and there is little scope for improvement
anyway).
In the harder sampling problem, where $g(\beta)$ was concave, the benefits of
tuning the $\{\beta_i\}$ are very clear.
In this example, the additional computational cost of tuning comes only from the \textsf{R} optimisation stage.
The benefits of tuning are greatest when $n$ is small (as $n$ increases,
the geometric $S_n$ anyway decreases to zero).

\subsubsection{When is the geometric temperature placement optimal?}

An interesting question raised by this example is under what circumstances will
tuning of the $\{\beta_i\}$ be likely to make efficiency gains
over the default geometric spacing for fixed $n$?
Suppose the target distribution is the $d$-dimensional multivariate Gaussian
with mean $\mu$ and variance $\Sigma$.
Then, $h(x) = \frac{1}{2} (x-\mu)^T \Sigma^{-1} (x-\mu)$ and the 
tempered distributions are $d$-dimensional multivariate Gaussian
with mean $\mu$ and variance $\beta_i^{-1}\Sigma$.
More importantly, $g(\beta)=\frac{d}{2\beta}$ and
$g^\prime(\beta) =\frac{-d}{2\beta^2}$.
As a result, when the $\{\beta_i\}$ are geometrically spaced, all the partial 
derivatives $\frac{ \partial S_n }{ \partial \beta_i}=0$ in Equation
(\ref{eqn:derivs2}), and so
geometric spacing is also the optimal minimum $S_n$ spacing.
In fact we can go further: suppose the set of
$\frac{ \partial S_n }{ \partial \beta_i}$ are all zero for a general $g$ and
for all $n$ when
the $\{\beta_i\}$ are geometrically spaced, i.e. when 
$\frac{\beta_{i+1}}{\beta_i}=c_n$ where
$c_n=\left ( \frac{\beta_n}{\beta_0} \right) ^{1/n}$, $\beta_n \ne 0$,
then
\begin{equation}
g^\prime (\beta_i) = - 
\frac{g\left (\frac{\beta_{i}}{{c_n}} \right )-2g(\beta_i)+g(c_n\beta_{i}) }
{ \frac{\beta_i}{c_n}-2\beta_i+c_n\beta_i} ,\ \ i=1,\ldots,n-1.
\label{eqn:jey1}
\end{equation}
As $n \rightarrow \infty$ with fixed $\beta_0$ and $\beta_n$, 
$c_n \rightarrow 1$ and a repeated application of L'h\^{o}pital's rule
yields the equation
\begin{equation}
\beta g^\prime (\beta) = - 
( \beta^2 g^{\prime\prime}(\beta) + \beta g^\prime( \beta))
\label{eqn:jey2}
\end{equation}
which has general solution $g(\beta) = \frac{K_1}{\beta} + K_2$ for
constants $K_1$ and $K_2$.
In other words, geometric spacing only minimises $S_n$ if the target
distribution has this form of $g(\beta)$.
This is a wider class than just the Gaussian, for example the 
exponential distribution has $g(\beta) = \frac{1}{\beta}$.
The result ties in with Figure \ref{fig:minsn} and Table \ref{tab:witch}.

\subsection{An alternative perspective on the optimisation problem}
\label{sec:alt_perspec}

In this section we provide an intuitive formalisation of what quantity 
$S_n$ in Equation (\ref{eqn:sums}) represents. 
Since 
$p_i(x)=\pi(x)\exp(-\beta_i h(x))/Z(\beta_i)$, it follows that
\begin{eqnarray*}
\frac{Z(\beta_{i+1})}{Z(\beta_i)} &=& \frac{p_i(x)}{\exp(-\beta_i h(x))}
\frac{\exp(-\beta_{i+1} h(x))}{p_{i+1}(x)} \nonumber \\
&=& \exp(-(\beta_{i+1}-\beta_i)h(x)) \frac{p_i(x)}{p_{i+1}(x)}
\end{eqnarray*}
Taking logarithms,
\begin{equation}
 \log\left( \frac{Z(\beta_{i+1})}{Z(\beta_i)} \right) = (\beta_i-\beta_{i+1})h(x)
- \log\left( \frac{p_{i+1}(x)}{p_i(x)} \right).
 \label{eqn:kl}
\end{equation}
Multiplying both sides of the equation by $p_{i+1}(x)$ and integrating 
with respect to $p_{i+1}(x)$ leads to
\begin{equation}
 \log\left( \frac{Z(\beta_{i+1})}{Z(\beta_i)} \right) = 
(\beta_i-\beta_{i+1})\mathsf{E}_{i+1}[h(X)] - KL\left[p_{i+1},p_i\right],
 \label{eqn:sum1}
\end{equation}
where $KL[p_{i+1},p_i] = \int_x p_{i+1}(x) \log\left( \frac{p_{i+1}(x)}{p_i(x)} \right)$,
is the Kullback-Leibler divergence between distributions $p_{i+1}$ and $p_i$. 
Similarly, it can be shown, by multiplying both sides of Equation (\ref{eqn:kl}) by $p_i(x)$ and 
integrating with respect to $x$, that
\begin{equation}
 \log\left( \frac{Z(\beta_{i+1})}{Z(\beta_i)} \right) = 
(\beta_i-\beta_{i+1})\mathsf{E}_i [h(X)] + KL[p_i,p_{i+1}].
 \label{eqn:sum2}
\end{equation}
Summing both Equations (\ref{eqn:sum1}) and (\ref{eqn:sum2}) over $i$ indices leads to
\begin{eqnarray}
 \log\left( \frac{Z(\beta_n)}{Z(\beta_0)} \right) 
&=& \sum_i(\beta_i-\beta_{i+1})\mathsf{E}_{i+1}[h(X)] - \sum_i KL[p_{i+1},p_i]
\nonumber \\
&=& \sum_i(\beta_i-\beta_{i+1})\mathsf{E}_{i}[h(X)] + \sum_i KL[p_{i},p_{i+1}]. \nonumber
\end{eqnarray}
It now follows directly that
\begin{displaymath}
 S_n(\beta_0,\ldots,\beta_n) = \frac{1}{2} \sum_{i=0}^{n-1} 
 \left\{ KL[p_{i+1},p_i]+ KL[p_{i},p_{i+1}] \right\}
\end{displaymath}
Thus our optimisation problem can be recast as one of finding temperatures
$\{\beta_1,\dots,\beta_{n-1}\}$ to minimise the sum of the symmetrised 
Kullback-Leibler distances between
successive distributions $p_i$ and $p_{i+1}$. 
This interpretation ties in with the recent work by Lefebvre, Steele and
Vandal (2010)\nocite{Lef08} who consider this same symmetrised 
Kullback-Leibler divergence in picking optimal schemes for path sampling.
A similar perspective, but in the context of marginal likelihood estimation using the 
power posterior method of Friel and Pettitt (2008)\nocite{Fri08}, appears in Section 3.2 of
that paper and also in Calderhead and Girolami (2009)\nocite{Calder08}. 

\section{Application of the tuning to a non-toy problem}
\label{sec:applications}

\subsection{A Bayesian mixture problem}
We now turn to Bayesian mixture modelling, an application where tempered 
transitions has been advocated in the past as a possible solution to 
sampling problems (see, for example,
Celeux, Hurn and Robert (2000)\nocite{CeleuxComputational} and
Jasra, Holmes and Stephens (2005)\nocite{Jas05}).
The benchmark for good MCMC mixing here is label-switching:
In the Bayesian treatment of a $k-$component mixture model, the likelihood 
is invariant to the labelling of the components.
This invariance is inherited by the posterior if, as is quite natural in
many cases, the priors do not impose identifiability.
The logical conclusion of such invariance is that a well-mixing MCMC 
sampler should visit all $k!$ labellings of the components.
Label-switching could be achieved trivially by incorporating a move type
which permutes the component labels, however this may mask more significant
difficulties in moving around the state space.
Certainly we can have greater confidence in the exploratory powers of
a sampler which can swap component labels in the course of its other moves.

\begin{figure}[t] 
\begin{center}
\includegraphics[height=12.5cm, angle=270]{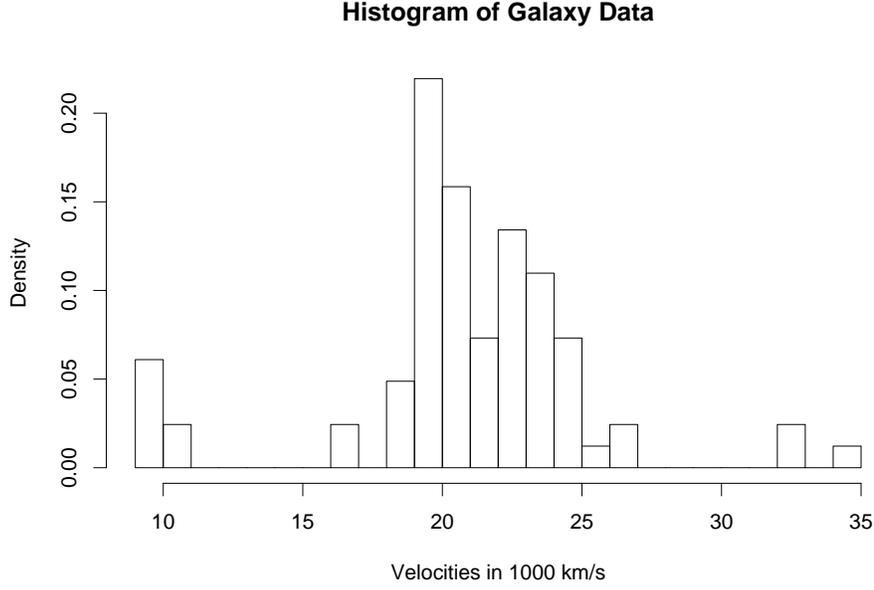}
\caption{
The galaxy data used for as illustration of the mixture modelling.}
\label{fig:gal}
\end{center}
\end{figure}

We use the much-studied galaxy data set for illustration, see for 
example Richardson and Green (1997)\nocite{RichardsonBayesian}, which comprises
measurements on the velocities of 82 galaxies (Figure \ref{fig:gal}).
Unlike in that paper though, we fix the number of mixture components at
three, the smallest number of components with non-negligible posterior
probability according to Richardson and Green.
Using a small number of components makes label switching 
harder as there is no ``redundant'' component to move freely
around the state space exchanging identities with the less mobile components 
needed to explain the data if they pass sufficiently close.

Denoting the 82 velocity measurements by $y = \{y_1,\ldots,y_{82}\}$, 
we follow Richardson and Green (1997)\nocite{RichardsonBayesian}
in incorporating corresponding latent allocation variables 
$z = \{z_1,\ldots,z_{82}\}$.
Given $z_i=j$, $y_i$ follows the $j^{th}$ of the three component Gaussian 
distributions of the mixture,
\begin{displaymath}
f( y_i | z_i=j , \mu_j,\sigma^2_j ) = \frac{1}
{\sqrt{2 \pi \sigma^2_j}} \exp
\left ( \frac{-(y_i-\mu_j)^2}{2 \sigma^2_j} \right ) \ \ \ i=1,\ldots,82.
\end{displaymath}
Further, conditional independence is assumed for the observations.
We specify largely independent standard proper priors:
\begin{eqnarray}
\mathsf{P}( z_i = j ) & = & w_j,\ \ \ \mbox{where } \sum_{j=1}^3 w_j = 1 \nonumber \\
\{ w_1 , w_2 , w_3 \} & \sim & Dirichlet(1,1,1) \nonumber \\
\mu_j & \sim & N( 0 , 1000 ), \ \ \ j=1, 2, 3 \nonumber \\
\sigma^2_j & \sim & InvGam(1,1), \ \ \ j=1, 2, 3. \nonumber
\label{eqn:3prior}
\end{eqnarray}
so that the posterior of interest is
\begin{equation}
f( z , \{w_j,\mu_j,\sigma^2_j\}_{j=1}^3 | y ) \propto 
\prod_{i=1}^{82} f( y_i | z_i , \mu_{z_i},\sigma^2_{z_i} ) \times
f( \{w_j\} ) \times \prod_{j=1}^3 f( \mu_j)\times
\prod_{j=1}^3 f( \sigma^2_j ) \times
\prod_{i=1}^{82} f( z_i | \{w_j\} )
\label{eqn:post}
\end{equation}

\begin{figure}[t] 
\begin{center}
\includegraphics[width=13cm,height=13cm]{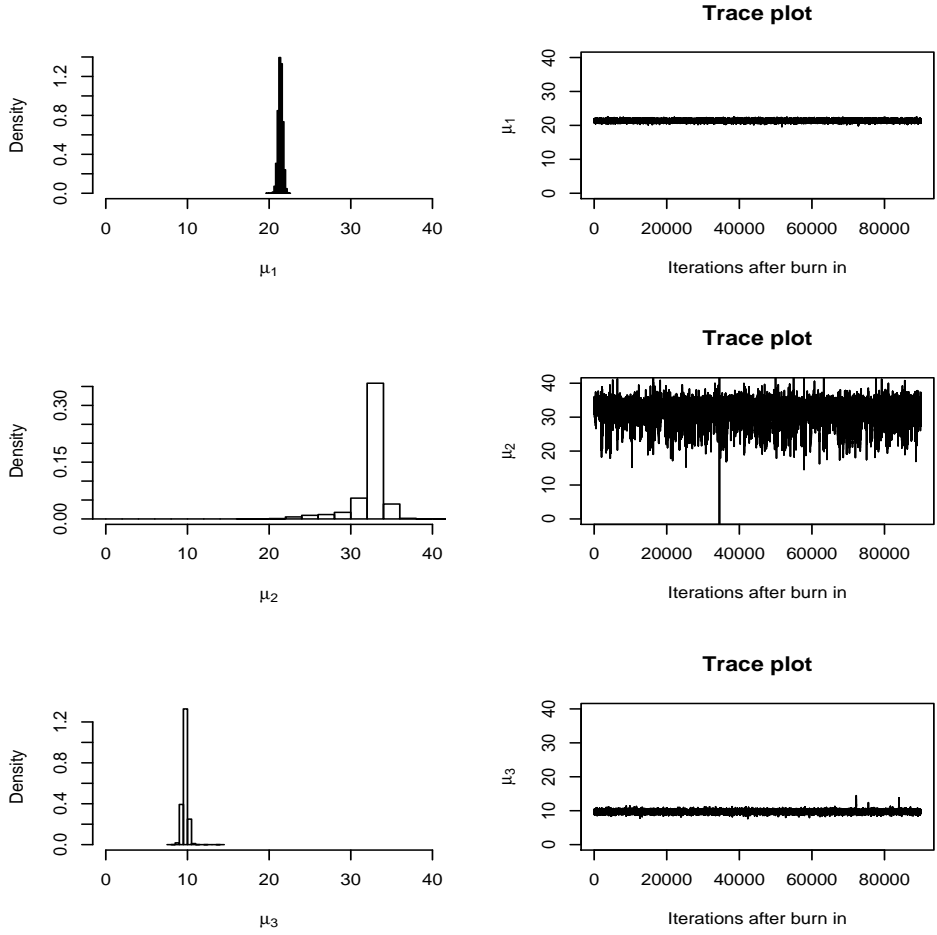}
\caption{
Histograms and trace plots of the $\{\mu_j\}$ chains indicating a lack of
mode swapping when $\beta=1$.}
\label{fig:nomix}
\end{center}
\end{figure}

We know that if label switching is taking place when sampling from
Equation~(\ref{eqn:post}) that the marginal posterior distributions for the 
sets of parameters of the three Gaussian components should be identical.
Figure~\ref{fig:nomix} shows the output for the $\{\mu_j\}$ parameters
using 100000 iterations of standard MCMC updates including a burn in of 10000
iterations (Gibbs updates for 
$\{w_j\},\{\mu_j\},\{\sigma^2_j\}$ and a uniform Metropolis proposal 
to change the $\{z_i\}$ in turn); it is clear that label switching is not
happening.
Tempering the whole of the posterior defined by Equation~(\ref{eqn:post})
is problematic as there is no guarantee that the
tempered distributions will remain proper.
Instead, we follow Celeux, Hurn and Robert (2000)\nocite{CeleuxComputational}
in tempering only the likelihood contribution leaving the priors
untempered.
This approach generates proper tempered distributions provided the priors are proper.
In the notation of previous sections, we set $\beta_0=1$ and
$\beta_n=1/16$ while
\begin{displaymath}
h( x) = \sum _{j=1}^3 \left ( \frac{n_j}{2}\log(\sigma_j^2) + 
\frac{1}{2\sigma^2_{j}} \sum_{\stackrel{i=1}{z_i = j}}^{82} ( y_i - \mu_{j} )^2 
\right )
\end{displaymath}
where $x=\{ z,\{w_j,\mu_j,\sigma^2_j\}_{j=1}^3\}$ and
$n_j = \sum_{i=1}^{82} I_{[z_i = j]},\ j=1,2,3$.
Unlike in the motivating example of the previous section, the $g(\beta)$
corresponding to this form of $h(x)$ is not available analytically and
so we must now address the question of approximating it before we can optimise
the $\{\beta_i\}$.

\subsection{Approximating $g(\beta)$}

The difficulty in estimating $g(\beta)=\mathsf{E}_{\beta}[ h(X)]$ and
$g^\prime(\beta)=-(\mathsf{E}_{\beta}[ h(X)^2]-\mathsf{E}_{\beta}[ h(X)]^2)$ for $\beta_n \le \beta \le \beta_0$ is that sampling
under $p_\beta$ is difficult for $\beta$ close to $\beta_0$ (hence the
need for tempered transitions!).
We propose instead to estimate $g(\beta)$ and
$g^\prime(\beta)$ using
importance sampling.
The obvious importance distribution to use is $p_{\beta_n}$ since we
have already made an assumption that we can sample from this distribution quite freely.
However it
may be a poor choice as an importance distribution for $p_\beta$ when
$\beta$ is close to $\beta_0$ because
when the importance distribution is quite far from the target,
the resulting estimates can be dominated by a handful of the 
samples (Robert and Casella 1999\nocite{RobertMC}).
As a compromise, we importance sample for expectations under $p_\beta$ 
by sampling under $p_{\tilde{\beta}}$ for
some $\tilde{\beta}$ where $\beta_n \le \tilde{\beta} < \beta \le \beta_0$,
in which case the unnormalised importance weights are $\exp( -(\tilde{\beta} - \beta)h(x))$.
We note in passing that a standard result states that the 
importance distribution which minimises the variance of the importance estimate
of some function $\psi(x)$ is
\begin{displaymath}
f ^* (x) \propto | \psi(x) | \ p_{\beta}(x).
\end{displaymath}
We turn this statement around to ask for what function $\psi(x)$ is
$p_{\tilde{\beta}}(x)$ the optimal importance distribution?
\begin{eqnarray*}
|\psi(x) | & \propto & \frac{p_{\tilde{\beta}}(x)}{p_{\beta}(x)} \nonumber \\
& = & \exp( -( \tilde{\beta} - \beta ) h(x) ) \nonumber \\
& = & 1 -( \tilde{\beta} - \beta ) h(x) + (( \tilde{\beta} - \beta ) h(x))^2/2 + \ldots
\end{eqnarray*}
So using $\tilde{p_{\beta}}$ as an importance distribution would be optimal if we were trying to estimate $\exp( -( \tilde{\beta} - \beta ) h(x) )$.
It is not optimal for estimating $\mathsf{E}_{\beta}[h(X)]$ and $\mathsf{E}_{\beta}[h(X)^2]$,
however it may be more reasonable for this goal when $( \tilde{\beta} - \beta )$ is small, than if we were, say, trying to 
estimate $\mathsf{E}_{\beta}[X]$ or $\mathsf{E}_{\beta}[X^{2}]$.

\begin{figure}[t] 
\begin{center}
\includegraphics[height=11.5cm]{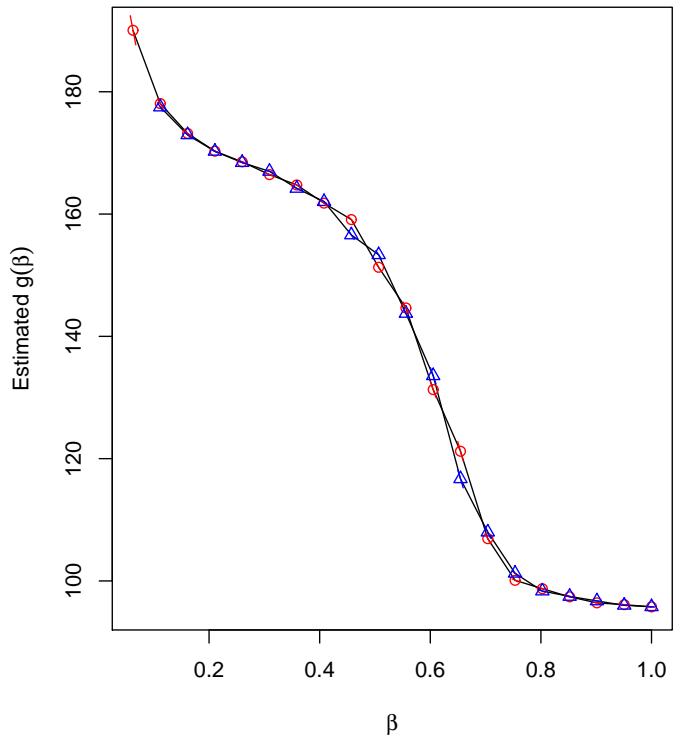}
\caption{
Estimating $g(\beta)$ and $g^\prime(\beta)$:
Symbols indicate the $\tilde{\beta}$ at which samples are generated;
red circles and red lines segments indicate direct sampling;
blue triangles and blue line segments indicate importance sampling
estimates; 
black lines are linear interpolations.}
\label{fig:robust}
\end{center}
\end{figure}

We work with 20 uniformly spaced values of $\tilde{\beta}$ in the interval
$[\beta_n,\beta_0]$. 
As a compromise between the inadequate sampling for large $\beta$ and the 
risk of unreliable importance sampling for large $\beta-\tilde{\beta}$,
we generate relatively small samples at each $\tilde{\beta}$
and use these samples to estimate $g(\tilde{\beta})$ and
$g^\prime(\tilde{\beta})$ both directly and indirectly by importance
sampling using the next smallest of the 20 chosen values (with
obvious modifications at the end points).
Figure \ref{fig:robust} shows the results when
using 10000 samples at each $\tilde{\beta}$
and discarding the first 1000 iterations as burn in.
We propose to use the average of the two estimates for $g(\tilde{\beta})$ and
$g^\prime(\tilde{\beta})$ at each point, with visual inspection recommended to 
check for major discrepancies.
In this example, the estimated $g(\beta)$ curve appears quite far from
 the geometric-friendly form $g(\beta)=\frac{K_1}{\beta} + K_2$.

\subsection{Results}

\begin{figure}[t] 
\begin{center}
\includegraphics[width=14.5cm, height=10.5cm]{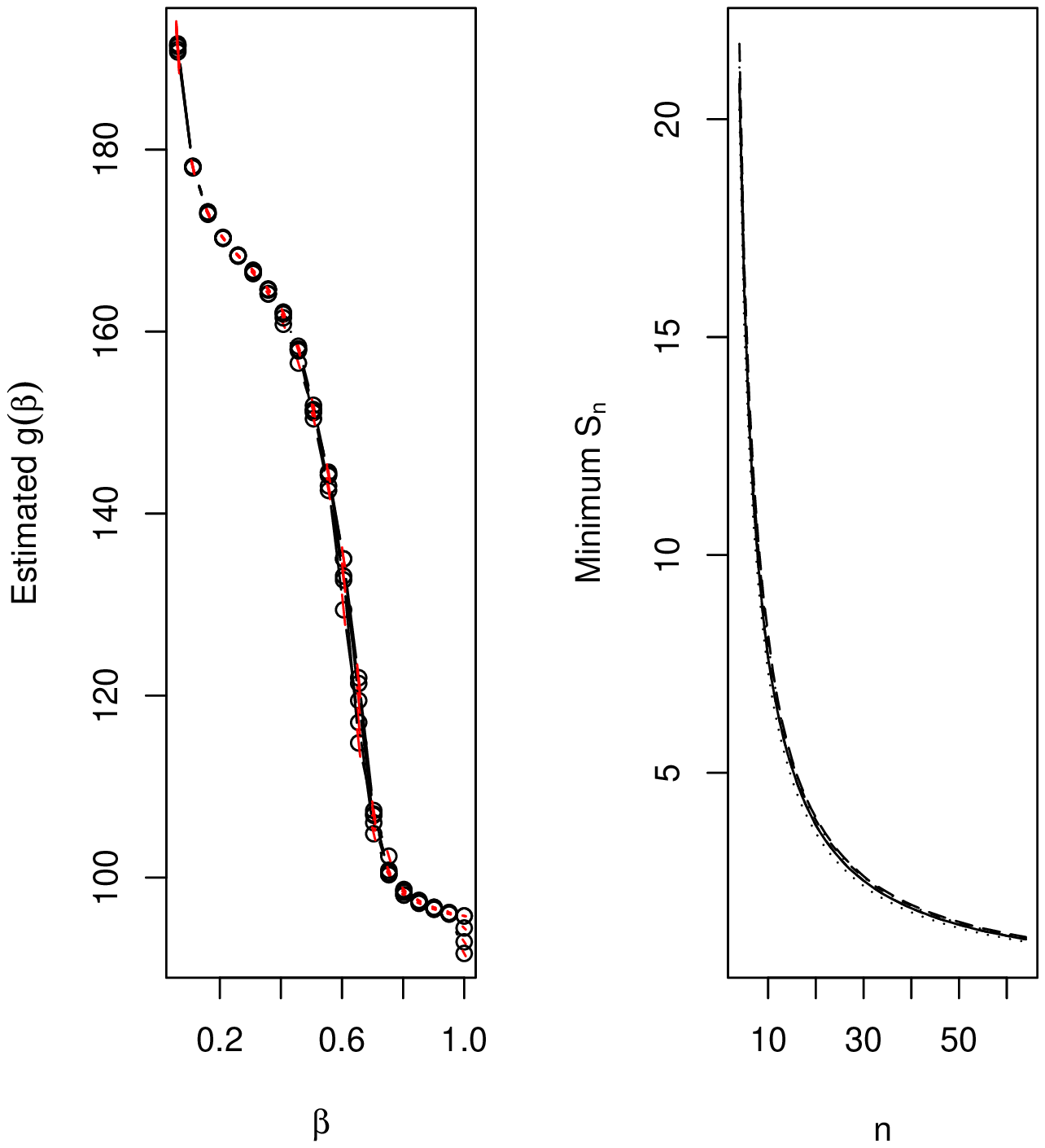}
\caption{Left: five replicates of the estimated $g(\beta)$ in black and 
$g^\prime(\beta)$ in red line segments with $\{\tilde{\beta}\}$ indicated
by circles.
Right: the five corresponding minimum $S_n$ against the number of tempering
levels $n$.}
\label{fig:minsum}
\end{center}
\end{figure}

Given the importance sampling estimates of both curve $g(\beta)$ and its 
derivative $g^\prime(\beta)$, we can
minimise $S_n(\beta_0,\ldots,\beta_n)$ using Equation (\ref{eqn:derivs2}).
As before, we use the \textsf{R} optimisation routine \textsf{optim} with linear
interpolation used to evaluate $g$ and $g^\prime$ between the 20
$\tilde{\beta}$ values.
In order to assess the effects of the imperfect estimation of $g(\beta)$
on the tuning procedure, we replicate the estimation process five times
with each replicate being used to select $\{\beta_i\}$.
Figure \ref{fig:minsum} shows both the variability in estimated $g$ and 
$g^\prime$ and how the optimised $S_n$ decreases with $n$ for 
the five sets of estimates.
By letting $n$ become sufficiently large, it would be possible to reduce
$S_n$ below any positive threshold.
(An upper bound on the minimum $S_n$ is 
$\frac{1}{n}(\beta_0-\beta_n)(g(\beta_n)-
g(\beta_0))$, achieved either by uniformly spacing 
$\beta_1,\ldots,\beta_{n-1}$ or by uniformly spacing 
$g(\beta_1),\ldots,g(\beta_{n-1})$.)
However as the computational cost of the tempering increases
linearly in $n$, the curves show that costs grow quite rapidly for
relatively small decreases in $S_n$.

Turning to the tempered transitions themselves, we ran the algorithm
for 100000 iterations including a burn-in of 10000 iterations.
At each tempering stage, the same proposal types were used as in the importance
sampling.
For each of the five sets of importance estimates, we temper using $n= 
64, 128, 256, 512$ 
and compare the optimised $\beta$ results with those of geometric
spacing.
We know that if switching is taking place, then the marginals
for each component should be identical.
This obviously implies that the three posterior expected $\mu_i$ should be
equal, as should the expected weights and the expected variances.
We propose to monitor the mixing using the usual tool of the integrated
autocorrelation times, however in estimating these diagnostics we use 
the averages of each group of parameters over the three chains rather than
the usual chain-wise average for each parameter.
For example, if the labels swap regularly, the individual averages of each of
the three $\mu_i$ chains will be close to their overall average, and
the non-centred autocorrelations will not be much different from the
standard centred ones.
On the other hand, if the labels do not switch as was the case with standard 
MCMC illustrated in Figure \ref{fig:nomix}, the autocorrelations 
calculated around the overall average will be greatly increased and this will 
be reflected in the modified integrated autocorrelation times.

\begin{table}[tbp]
\begin{center}
\begin{tabular}{|l|c|c|c|c|}
\hline
 & $n=64$ & $n=128$ & $n=256$ & $n=512$ \\
\hline
Geometric & $\alpha$=0.00013 & $\alpha$=0.00065 & $\alpha$=0.00187 & $\alpha$=0.00853 \\
$\hat{\tau}(\{w_j\})$ & ***, ***, *** & 3253, 22910, 95995 & 1649, 1382, 1916 & 256, 249, 257 \\
$\hat{\tau}(\{\mu_j\})$ & 96042, ***, *** & 4505, 6938, 4552 & 1003, 1457, 745 &322, 322, 269 \\
$\hat{\tau}(\{\sigma^2_j\})$ & ***, 4970 5188 & 477, 768, 1137 & 710, 475, 799 & 140, 132, 163 \\
\hline
Tuned 1 & $\alpha$=0.00062 & $\alpha$=0.00316 & $\alpha$=0.01366 & $\alpha$=0.04105 \\
$\hat{\tau}(\{w_j\})$ & 9089, 4750, 5743 & 1137, 721, 1117 & 161, 148, 161 & 50, 51, 45 \\
$\hat{\tau}(\{\mu_j\})$ & 6406, 3842, 2702 & 781, 762, 952 & 134, 165, 160 & 45, 48, 49 \\
$\hat{\tau}(\{\sigma^2_j\})$ & 3071, 1323, 5498 & 321, 113, 441 & 78, 88, 90 & 31, 33, 28 \\
\hline
Tuned 2 & $\alpha$=0.00054 & $\alpha$=0.00346 & $\alpha$=0.01426 & $\alpha$=0.04194 \\
$\hat{\tau}(\{w_j\})$ & 5481, 10701, 17267 & 642, 1357, 952 & 134, 161, 148 & 51, 45, 49 \\
$\hat{\tau}(\{\mu_j\})$ & 13077, 6291, 20823 & 893, 1235, 1309 & 148, 123, 133 & 48, 54, 51 \\
$\hat{\tau}(\{\sigma^2_j\})$ & 6040, 8845, 2274 & 249, 250, 239 & 116, 72, 44 & 38, 34, 26 \\
\hline
Tuned 3 & $\alpha$=0.00053 & $\alpha$=0.00362 & $\alpha$=0.01395 & $\alpha$=0.04467 \\
$\hat{\tau}(\{w_j\})$ & 8610, 4050, 3609 & 774, 1244, 813 & 185, 166, 178 & 49, 50, 49 \\
$\hat{\tau}(\{\mu_j\})$ & 6826, 19664, 25232 & 1499, 827, 1133 & 185, 167, 149 & 46, 47, 51 \\
$\hat{\tau}(\{\sigma^2_j\})$ & 5967, 2569, 2283 & 268, 288, 156 & 83, 91, 56 & 34, 37, 23\\
\hline
Tuned 4 & $\alpha$=0.00054 & $\alpha$=0.00282 & $\alpha$=0.00923 & $\alpha$=0.03884 \\
$\hat{\tau}(\{w_j\})$ & 4668, 8875, 8538 & 960, 1043, 862 & 258, 253, 244 & 51, 50, 55 \\
$\hat{\tau}(\{\mu_j\})$ & 6335, 3805, 16416 & 2365, 751, 1928 & 250, 197, 209 & 51, 47, 53 \\
$\hat{\tau}(\{\sigma^2_j\})$ & 3266, 4463, 1782 & 411, 218, 376 & 91, 115, 117 & 43, 32, 39 \\
\hline
Tuned 5 & $\alpha$=0.00062 & $\alpha$=0.00275 & $\alpha$=0.00928 & $\alpha$=0.02518 \\
$\hat{\tau}(\{w_j\})$ & 18057, 6343, 7629 & 1065, 763, 1353 & 293, 261, 210 & 75, 77, 86 \\
$\hat{\tau}(\{\mu_j\})$ & 5513, 5965, 39982 & 3743, 1671, 1966 & 221, 272, 287 & 91, 93, 87 \\
$\hat{\tau}(\{\sigma^2_j\})$ & 1423, 1496, 538 & 492, 506, 386 & 118, 149, 118 & 59, 48, 64 \\
\hline
\end{tabular}
\caption{Results for the mixture problem using different
numbers of tempering levels. 
Results are shown for geometric spacing of the $\{\beta_i\}$
and for the tuned spacing from the five replicates of importance
sampling.
*** indicates that the estimates of integrated autocorrelation times
did not converge reliably.}
\label{tab:tuned}
\end{center}
\end{table}

Table \ref{tab:tuned} summarises the results, showing the acceptance rates and
the estimated integrated autocorrelations times for the three sets of
parameters.
The first point to note is how large $n$ needs to be in order to achieve
even low acceptance rates.
This is not unexpected; Jasra, Holmes and Stephens (2005)\nocite{Jas05}
describe ``huge rejection rates when sampling from the full posterior'' using
tempered transitions. 
Although tempering moves may not be accepted very often, each one can 
make a large move in the state space and it is common practice to intersperse
tempering moves with standard MCMC moves for improved local exploration.
The fact that acceptance rates can be so low highlights the importance of
any tuning.
The worst case is geometric spacing when $n=64$; here the actual number of acceptances
is so low, just 13, that the estimates of integrated autocorrelation times fail to
converge reliably (taken to mean that the estimate exceeded a tenth of
the total run length).
As $n$ increases, acceptance rates improve and integrated autocorrelation
times decrease for all runs.
There is some variability between the five replicates of the tuned spacings
of the $\{\beta_i\}$, however at all of the $n$ considered, all five outperform geometric spacing by some
considerable margin in terms of the acceptance rates and consequently the
integrated autocorrelation times. 

The cost of tuning the $\{ \beta_i \}$ comprises the cost of the samples 
required to estimate $g(\beta)$ and $g^\prime(\beta)$ using importance sampling 
plus the optimisation costs for minimising $S_n$.
Here we used 20 relatively short runs, each of only 10000 iterations, for the 
$g(\beta)$ and $g^\prime(\beta)$ estimation; this stage is independent
of $n$.
The cost of the deterministic minimisation of $S_n$ 
in \textsf{R} increases with $n$ but is of the order of a few minutes for $n=512$. 
This makes the cost of the tuning procedure a small fraction of the total cost. 
Full details are given in Table \ref{table:time} showing
that the additional cost of the tuning procedure, for this example, varied
between $2\%$ and $4\%$ extra CPU time compared to the untuned procedure.
Combining this information with the integrated autocorrelation times in Table~\ref{tab:tuned} indicates
the tuned procedure gives substantial improvements in mixing compared to the geometric temperature
placement.

\begin{table}[tbp]
\begin{center}
\begin{tabular}{|l|r|r|r|r|}
\hline
 & $n=64$ & $n=128$ & $n=256$ & $n=512$ \\
\hline
Estimating $g$ and $g^\prime$ & 8.35  & 8.35  & 8.35  & 8.35  \\
Optimisation & 2.78 & 16.97 & 52.02 & 191.00 \\
Tuned tempering & 879.81 & 1762.61 & 3507.79 & 7022.67  \\ \hline
Tuned total & 890.94 & 1787.93 & 3568.16  & 7222.02 \\
\hline
Geometric tempering & 874.12 & 1758.88 & 3494.94 & 7002.63 \\
\hline
Proportion increase & 1.019 & 1.017 & 1.021 & 1.031 \\
\hline
\end{tabular}
\caption{Time in user CPU time seconds for the geometrically spaced temperatures and for the tuned temperatures including a breakdown of the cost of tuning.}
\label{table:time}
\end{center}
\end{table}

In this example, very little of the total computational effort was spent in
estimating $g(\beta)$ and $g^\prime(\beta)$.
Although importance sampling cannot be guaranteed to be particularly good for this type of problem, we suggest that this is a sensible strategy.
The associated risk is either that the importance sampling fails to identify
a $g(\beta)$ curve which is not suitable for geometric spacing or,
conversely,
that it identifies interesting features which are not in fact present.
In the former case, a visual inspection
of the roughly estimated $g(\beta)$ may suggest that it is not
worthwhile implementing any optimisation, reverting to the default
geometric, and so the wasted CPU time is minimal.
(The same argument is also reasonable when importance sampling 
works well for estimating $g(\beta)$.)
On the other hand, if the estimated $g(\beta)$ looks to be of the form where tuning may help, more computational effort could be put into improving the accuracy of its estimation, especially if a discrepancy is
noted between the values of the curve using direct and indirect sampling.

\section{Discussion}
\label{sec:discussion}
In this paper we have explored how to tune the expensive tempered
transitions algorithm to make best use of computational resources.
We have shown that the geometric schedule will be optimal if the curve
$g(\beta)=\mathsf{E}_\beta [h(X)]$ is of a particular form, where the target
distribution is $p(x) \propto \pi(x)\, \exp( -\beta_0 \, h(x) )$.
The tuning itself is relatively cheap and examples have demonstrated that it can
make a significant difference.

Although we have not explicitly considered the question of choosing
the number of tempering levels, we have some purely anecdotal evidence
that the tuning procedure may yield 
useful information regarding the minimum number of tempering levels required.
In our experience, tempered transitions does not seem to perform at all well with 
a $\{\beta_i\}$ for which $S_n > 1$.
For example, in our mixture example the geometric $S_n$ is approximately of the order 
2, 1, 0.5 and 0.25 for
$n=64, 128, 256, 512$ respectively, while for the tuned $\{\beta_i\}$
it is of the corresponding approximate value 1.2, 0.6, 0.3 and 0.15.
It also seems feasible that the tuning approach proposed here may also
be relevant to some of the other MCMC algorithms which incorporate an
element of tempering.
This is another topic for future research.

\section*{Acknowledgements}
Gundula Behrens thanks the Engineering and
Physical Sciences Research Council and Evangelisches Studienwerk for
financial support.
Nial Friel's research was supported by a visit to Bath by the Bath Institute for Complex 
Systems (EPSRC grant GR/S86525/01) and by a Science Foundation
Ireland Research Frontiers Program grant, 09/RFP/MTH2199.
We are grateful to the Associate Editor and the referees for their 
insightful and helpful comments and to Dr Jey Sivaloganathan for 
advice on Equations (\ref{eqn:jey1}) and (\ref{eqn:jey2}).

\bibliographystyle{plain}
\bibliography{tuningPaperBibliography}

\end{document}